\documentclass[twocolumn,showpacs,preprintnumbers,amsmath,amssymb,prb]{revtex4}

\usepackage{graphicx}
\usepackage{dcolumn}
\usepackage{bm}
\usepackage{tabularx}
\usepackage{amsmath}

\begin{document}

\title{Thermodynamic properties of Ba$_{1-x}$M$_x$Fe$_2$As$_2$ (M = La and K)}

\author{J. K. Dong,$^1$ L. Ding,$^1$ H. Wang,$^1$ X. F. Wang,$^2$ T. Wu,$^2$ X. H. Chen,$^2$ and S. Y. Li$^{1,*}$}

\affiliation{$^1$Department of Physics, Surface Physics Laboratory (National Key Laboratory), and Advanced Materials Laboratory, Fudan University, Shanghai 200433, P. R. China\\
$^2$Hefei National Laboratory for Physical Science at Microscale and
Department of Physics, University of Science and Technology of
China, Hefei, Anhui 230026, P. R. China}

\date{\today}

\begin{abstract}
The specific heat $C(T)$ of BaFe$_2$As$_2$ single crystal,
electron-doped Ba$_{0.7}$La$_{0.3}$Fe$_2$As$_2$ and hole-doped
Ba$_{0.5}$K$_{0.5}$Fe$_2$As$_2$ polycrystals were measured. For
undoped BaFe$_2$As$_2$ single crystal, a very sharp specific heat
peak was observed at 136 K. This is attributed to the structural and
antiferromagnetic transitions occurring at the same temperature.
$C(T)$ of the electron-doped non-superconducting
Ba$_{0.7}$La$_{0.3}$Fe$_2$As$_2$ also shows a small peak at 120 K,
indicating a similar but weaker structural/antiferromagnetic
transition. For the hole-doped superconducting
Ba$_{0.5}$K$_{0.5}$Fe$_2$As$_2$, a clear peak of $C/T$ was observed
at $T_c$ = 36 K, which is the highest peak seen at superconducting
transition for iron-based high-$T_c$ superconductors so far. The
electronic specific heat coefficient $\gamma$ and Debye temperature
$\Theta_D$ of these compounds were obtained from the low temperature
data.

\end{abstract}


\maketitle

The recent discovery of superconductivity with $T_c$ as high as 55 K
in iron-based LnO$_{1-x}$F$_x$FeAs (Ln = La, Sm, Ce, Nd, Pr, Gd, Tb,
Dy) has attracted great attention.
\cite{Kamihara,XHChen1,ZARen1,XHChen2,GFChen1,ZARen2,ZARen3,HHWen1,Bos}
It is believed that the FeAs layers are responsible for the
high-$T_c$ superconductivity and the LnO layers provide electron
carriers through fluorine
doping,\cite{Kamihara,XHChen1,ZARen1,XHChen2,GFChen1,ZARen2,ZARen3,HHWen1,Bos}
by simply introducing oxygen vacancies, \cite{ZARen4} or by
substituting Ln$^{3+}$ with Th$^{4+}$. \cite{CWang} This is very
similar to the high-$T_c$ cuprate superconductors. Since increasing
the number of CuO$_2$ layers in one unit cell has been a promising
way to elevate $T_c$ for cuprates, many efforts have been put into
searching superconductivity in iron-based compounds with multiple
FeAs layers.

The ternary iron arsenide BaFe$_2$As$_2$ with the tetragonal
ThCr$_2$Si$_2$-type structure was first suggested by Rotter {\it et
al.} \cite{Rotter1} that it can serve as a parent compound for
oxygen-free iron arsenide superconductors. Very soon,
superconductivity with $T_c$ = 38 K was indeed discovered in
Ba$_{1-x}$K$_x$Fe$_2$As$_2$. \cite{Rotter2} In contrast to
electron-doped LnO$_{1-x}$F$_x$FeAs, the carriers in
Ba$_{1-x}$K$_x$Fe$_2$As$_2$ are holes introduced by potassium
doping, which has been verified by Hall coefficient and
thermoelectric power measurements. \cite{GWu1} Later,
superconductivity was also found in Sr$_{1-x}$M$_x$Fe$_2$As$_2$ (M =
K and Cs) and Ca$_{1-x}$Na$_x$Fe$_2$As$_2$ with $T_c \sim$ 38 and 20
K, respectively. \cite{GFChen2,Sasmal,GWu2} Interestingly, the
electron-doped Ba$_{1-x}$La$_x$Fe$_2$As$_2$ shows no sign of
superconductivity. \cite{GWu1}

For the parent compound BaFe$_2$As$_2$, resistivity, specific heat,
and susceptibility show clear anomaly near 140 K.
\cite{Rotter1,XFWang} Similar anomalies were also observed in
SrFe$_2$As$_2$ and EuFe$_2$As$_2$ at relatively higher temperature
near 200 K. \cite{GFChen2,Krellner,ZhiRen,Jeevan} Neutron scattering
experiment has demonstrated that in BaFe$_2$As$_2$ there is a phase
transition to a long-ranged antiferromagnetic state at 142 K where a
first-order structural transition from tetragonal to orthorhombic
symmetry also occurs. \cite{QHuang} This is analogous to that in
LaOFeAs compound, where structural and spin-density-wave transitions
were observed, but occurring at different temperatures, 155 K and
137 K respectively. \cite{Dai} In both LnOFeAs and AFe$_2$As$_2$ (A
= Ba, Sr, and Ca), antiferromagnetism gives way to superconductivity
as electrons or holes are doped, indicating antiferromagnetic spin
fluctuations may play an important role in these systems.

While the specific heat peak near 140 K was observed in
BaFe$_2$As$_2$ polycrystalline sample, \cite{Rotter1} it was absent
from the $C(T)$ curve of the first reported BaFe$_2$As$_2$ single
crystal. \cite{Ni} The resistivity behavior of their single crystal
\cite{Ni} was also different from previous polycrystalline sample.
\cite{Rotter2} These unusual resistivity and specific heat behaviors
may result from the contamination of Sn flux in the crystal.
\cite{Ni} Therefore it is interesting to check the intrinsic
specific heat behavior of BaFe$_2$As$_2$ single crystal with high
purity. There are also no specific heat data for superconducting
A$_{1-x}$M$_x$Fe$_2$As$_2$ (A = Ba, Sr, and Ca; M = K, Cs, and Na)
compounds so far.

In this Brief Report, we present the first specific heat results of
high quality BaFe$_2$As$_2$ single crystal, electron-doped
Ba$_{0.7}$La$_{0.3}$Fe$_2$As$_2$ and hole-doped
Ba$_{0.5}$K$_{0.5}$Fe$_2$As$_2$ polycrystals. Very sharp specific
heat peak at 136 K was observed for our BaFe$_2$As$_2$ single
crystal, 3 times higher than that in previous polycrystalline
sample. A much smaller peak near 120 K was also observed for
Ba$_{0.7}$La$_{0.3}$Fe$_2$As$_2$, indicating a weak
structural/antiferromagnetic transition in this electron-doped
sample. A clear peak of $C/T$ shows up at $T_c$ = 36 K for the
hole-doped Ba$_{0.5}$K$_{0.5}$Fe$_2$As$_2$ superconducting sample.
By fitting the low temperature data, the electronic specific heat
coefficient $\gamma$ and Debye temperature $\Theta_D$ of these
compounds were obtained.

The single crystals of BaFe$_2$As$_2$ and polycrystalline samples
with nominal composition Ba$_{0.7}$La$_{0.3}$Fe$_2$As$_2$ and
Ba$_{0.5}$K$_{0.5}$Fe$_2$As$_2$ were prepared as in Ref. 14 and 18.
By employing self-flux method (FeAs as flux), our BaFe$_2$As$_2$
single crystals are free of contamination from other elements.
Resistivity were measured by the standard four-probe method.
Specific heat measurements were performed in a Quantum Design
physical property measurement system (PPMS) via the relaxation
method. Magnetic field $H$ = 8 T was applied for the
Ba$_{0.5}$K$_{0.5}$Fe$_2$As$_2$ superconducting sample.

Fig. 1 shows the resistivity of BaFe$_2$As$_2$ single crystal,
Ba$_{0.7}$La$_{0.3}$Fe$_2$As$_2$ and Ba$_{0.5}$K$_{0.5}$Fe$_2$As$_2$
polycrystals. Abrupt resistivity drop can be seen at $T_s =$ 134 K
for BaFe$_2$As$_2$ single crystal, consistent with previous reports.
\cite{Rotter1,GWu1} Similar resistivity drop was also observed for
Ba$_{0.7}$La$_{0.3}$Fe$_2$As$_2$ sample, but less abrupt and
shifting to lower temperature $T_s =$ 122 K. No superconducting
transition was observed at low temperature down to 5 K for the
electron-doped Ba$_{0.7}$La$_{0.3}$Fe$_2$As$_2$, similar to previous
Ba$_{0.85}$La$_{0.15}$Fe$_2$As$_2$ sample. \cite{GWu1} The
hole-doped Ba$_{0.5}$K$_{0.5}$Fe$_2$As$_2$ sample shows a sharp
superconducting transition starting at 38 K and reaching zero
resistivity at 34 K. We use the middle point of transition 36 K as
its $T_c$.

\begin{figure}
\includegraphics[clip,width=7.7cm]{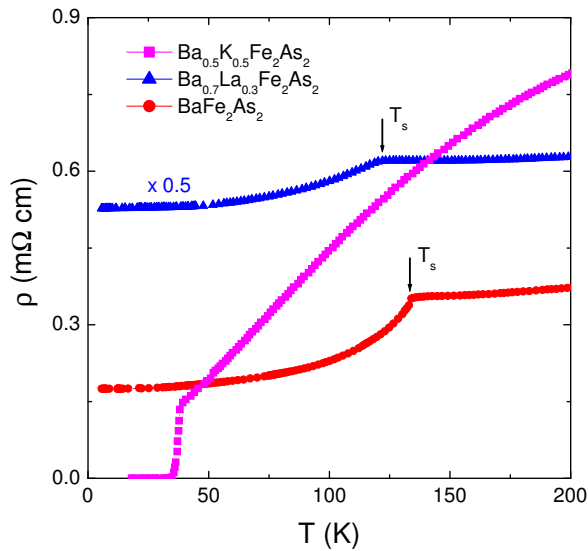}
\caption{(Color online) Resistivity of BaFe$_2$As$_2$ single
crystal, electron-doped Ba$_{0.7}$La$_{0.3}$Fe$_2$As$_2$ and
hole-doped Ba$_{0.5}$K$_{0.5}$Fe$_2$As$_2$ polycrystals. Arrows
denote the structural and antiferromagnetic transitions occurring at
the same temperature $T_s$.}
\end{figure}

The specific heat $C(T)$ of BaFe$_2$As$_2$,
Ba$_{0.7}$La$_{0.3}$Fe$_2$As$_2$ and Ba$_{0.5}$K$_{0.5}$Fe$_2$As$_2$
samples are shown in Fig. 2. For the BaFe$_2$As$_2$ single crystal,
one can see a very sharp peak (enlarged in the inset) with $\Delta C
\approx$ 130 J / mol K, which is different from the first reported
BaFe$_2$As$_2$ single crystal by Ni {\it et al.}, \cite{Ni} and
consistent with previous result on polycrystalline sample.
\cite{Rotter1} The height of specific heat peak in the insert of
Fig. 2 is also 3 times higher than that in polycrystalline sample
where $\Delta C \approx$ 35 J / mol K. \cite{Rotter1} This proves
the high quality of our BaFe$_2$As$_2$ single crystal. It is
believed that the use of self-flux method of growth gives the high
purity of our samples, thus the intrinsic specific heat behavior of
BaFe$_2$As$_2$ single crystal. As found in SrFe$_2$As$_2$,
\cite{Krellner} the abrupt resistivity drop and sharp specific heat
peak in Fig. 1 and 2 should be compatible with a first-order phase
transition in BaFe$_2$As$_2$. Neutron scattering experiment has
demonstrated such first-order structural phase transition,
accompanied by a long-range antiferromagnetic transition.
\cite{QHuang} The entropy connected with the transition $\Delta S
\sim$ 1 J / mol K is estimated from the area under the $C(T)$ peak
in the insert of Fig. 2. This small value of $\Delta S$ is almost
the same as that in SrFe$_2$As$_2$. \cite{Krellner}

In Fig. 2, $C(T)$ of the electron-doped
Ba$_{0.7}$La$_{0.3}$Fe$_2$As$_2$ sample also manifests a small peak
at 120 K. This is consistent with the resistivity drop at 122 K in
Fig. 1, indicating a similar but weaker structural/antiferromagnetic
transition. No anomaly of $C(T)$ was observed above $T_c$ for the
hole-doped Ba$_{0.5}$K$_{0.5}$Fe$_2$As$_2$ sample. These results
show that hole doping suppresses structural/antiferromagnetic
transition more efficiently than electron doping in
Ba$_{1-x}$M$_x$Fe$_2$As$_2$ system.

\begin{figure}
\includegraphics[clip,width=7.75cm]{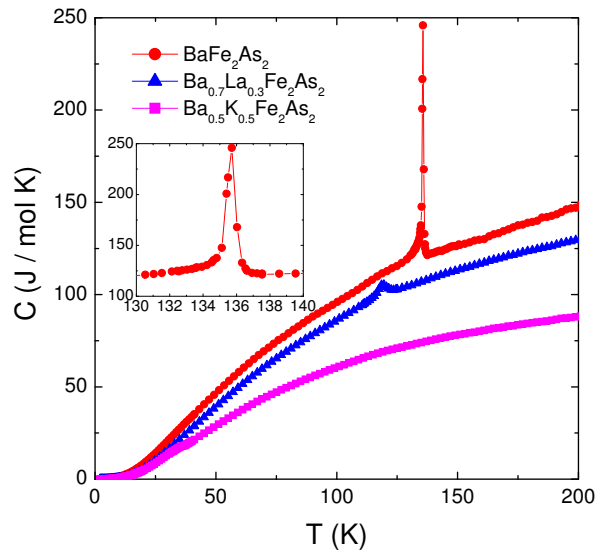}
\caption{(Color online) Specific heat of BaFe$_2$As$_2$,
Ba$_{0.7}$La$_{0.3}$Fe$_2$As$_2$ and Ba$_{0.5}$K$_{0.5}$Fe$_2$As$_2$
samples. The parent compound BaFe$_2$As$_2$ single crystal shows a
very sharp peak at 136 K, enlarged in the insert. A small peak is
also visible at 120 K for Ba$_{0.7}$La$_{0.3}$Fe$_2$As$_2$.}
\end{figure}

Fig. 3 plots $C/T$ vs $T$ for the superconducting
Ba$_{0.5}$K$_{0.5}$Fe$_2$As$_2$ sample from 30 to 40 K in zero and
$H$ = 8 T magnetic fields. A clear peak shows up at $T_c$ = 36 K and
it is suppressed to lower temperature by 8 T field. Rough estimation
gives the specific heat jump $\Delta C/T \approx$ 15 mJ / mol K$^2$
at $T_c$ in zero field. Previously, there was no visible anomaly on
$C/T$ near $T_c$ for LaO$_{1-x}$F$_x$FeAs, \cite{GMu,Sefat} and only
a small kink was observed at $T_c$ for SmO$_{1-x}$F$_x$FeAs.
\cite{Ding} To our knowledge, the $C/T$ peak in Fig. 3 is the
highest peak seen at the superconducting transition in iron-based
high-$T_c$ superconductors so far. This may be attributed to the
better quality, or higher superfluid density of the
Ba$_{0.5}$K$_{0.5}$Fe$_2$As$_2$ sample.

\begin{figure}
\includegraphics[clip,width=7.7cm]{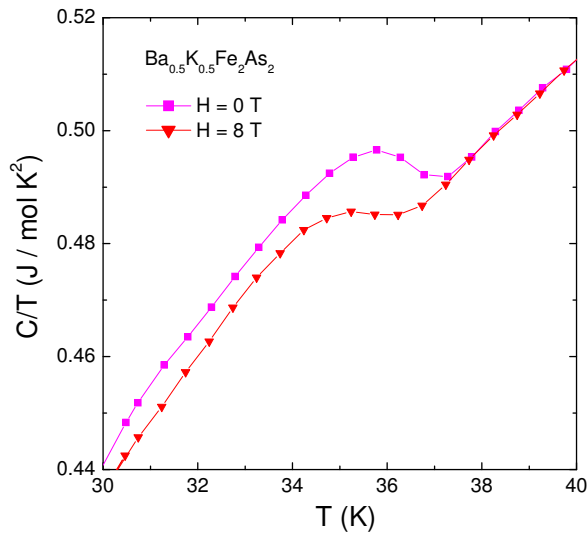}
\caption{(Color online) $C/T$ vs $T$ for
Ba$_{0.5}$K$_{0.5}$Fe$_2$As$_2$ polycrystalline sample near $T_c$ =
36 K in zero and $H$ = 8 T magnetic fields. Rough estimation gives
the specific heat jump $\Delta C/T \approx$ 15 mJ / mol K$^2$ at
$T_c$ in zero field.}
\end{figure}

\begin{figure}
\includegraphics[clip,width=7.7cm]{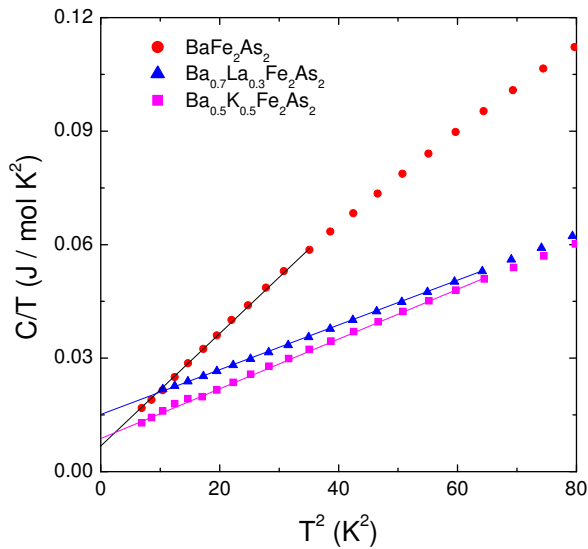}
\caption{(Color online) $C/T$ vs $T^2$ for BaFe$_2$As$_2$,
Ba$_{0.7}$La$_{0.3}$Fe$_2$As$_2$ and Ba$_{0.5}$K$_{0.5}$Fe$_2$As$_2$
samples. The lines are linear fits at low temperature.}
\end{figure}

In Fig. 4, $C/T$ vs $T^2$ is plotted for BaFe$_2$As$_2$,
Ba$_{0.7}$La$_{0.3}$Fe$_2$As$_2$ and Ba$_{0.5}$K$_{0.5}$Fe$_2$As$_2$
samples at low temperature. For BaFe$_2$As$_2$ single crystal, the
data between 2 and 6 K can be linearly fitted to $C/T = \gamma +
\beta T^2$, which gives the electronic specific heat coefficient
$\gamma$ = 6.1 $\pm$ 0.3 mJ / mol K$^2$ and $\beta$ = 1.51 $\pm$
0.01 mJ / mol K$^4$. This value of $\gamma$ is smaller than that
obtained in polycrystal sample (16 mJ / mol K$^2$), \cite{Rotter1}
and much smaller than that of the first reported BaFe$_2$As$_2$
single crystal (37 mJ / mol K$^2$). \cite{Ni} Using the equation
$\beta = (12\pi ^4 n k_B)/(5\Theta_D ^3)$, where $n$ is the number
of atoms per formula unit, we estimate the Debye temperature
$\Theta_D =$ 186 K for BaFe$_2$As$_2$ single crystal.

For Ba$_{0.7}$La$_{0.3}$Fe$_2$As$_2$ and
Ba$_{0.5}$K$_{0.5}$Fe$_2$As$_2$ samples, the data between 2 and 8 K
can also be fitted as above, which gives $\gamma$ = 15.2 $\pm$ 0.1
mJ / mol K$^2$, $\beta$ = 0.586 $\pm$ 0.003 mJ / mol K$^4$, and
$\gamma =$ 9.1 $\pm$ 0.2 mJ / mol K$^2$, $\beta$ = 0.653 $\pm$ 0.005
mJ / mol K$^4$, respectively. Comparing with the parent compound,
electron doping seems to increase $\gamma$ in the
non-superconducting Ba$_{0.7}$La$_{0.3}$Fe$_2$As$_2$. The residual
$\gamma =$ 9.1 mJ / mol K$^2$ of the superconducting
Ba$_{0.5}$K$_{0.5}$Fe$_2$As$_2$ sample suggests nodes in the
superconducting gap. However, since these two samples are
polycrystals, their values of $\gamma$ may be not completely
intrinsic. The Debye temperature of these two samples are estimated
to be $\Theta_D =$ 254 and 246 K, respectively.

In summary, we have studied the specific heat of high quality
BaFe$_2$As$_2$ single crystal, electron-doped
Ba$_{0.7}$La$_{0.3}$Fe$_2$As$_2$ and hole-doped
Ba$_{0.5}$K$_{0.5}$Fe$_2$As$_2$ polycrystals. For BaFe$_2$As$_2$
single crystal, a very sharp specific heat peak at 136 K was
observed, consistent with the polycrystal result. A small peak near
120 K was also observed for electron-doped
Ba$_{0.7}$La$_{0.3}$Fe$_2$As$_2$, indicating a weak
structural/antiferromagnetic transition. A clear peak of $C/T$ can
be seen at $T_c$ = 36 K for the hole-doped
Ba$_{0.5}$K$_{0.5}$Fe$_2$As$_2$ sample, which is the highest one
among all iron-based high-$T_c$ superconductors so far. By fitting
the low temperature data, we obtain the electronic specific heat
coefficient $\gamma$ and Debye temperature $\Theta_D$ of these
compounds.

This work is supported by the Natural Science Foundation of China,
the Ministry of Science and Technology of China (973 project No:
2006CB601001 and National Basic Research Program No:2006CB922005), and STCSM of China.\\

$^*$ Electronic address: shiyan$\_$li@fudan.edu.cn

\end{document}